\documentclass[proof]{WileyASNA-v1}

\articletype{Article Type}%

\received{30 September 2019}
\revised{17 February 2020}
\accepted{24 February 2020}

\begin{document}

\title{Dynamics of spherical space debris of different sizes falling to Earth}

\author[1,2]{Judit Sl\'{i}z-Balogh}

\author[2]{D\'aniel Horv\'ath}

\author[2]{R\'obert Szab\'o }

\author[2]{G\'abor Horv\'ath* }

\authormark{Judit Sl\'{i}z-Balogh \textsc{et al}}

\address[1]{\orgdiv{Department of Astronomy}, \orgname{E\"otv\"os Lor\'and University}, \orgaddress{\state{}\country{Hungary}}}

\address[2]{\orgdiv{Department of Biological Physics}, \orgname{E\"otv\"os Lor\'and University}, \orgaddress{\state{}\country{Hungary}}}

\corres{*G\'abor Horv\'ath, H-1117 Budapest, P\'azm\'any s\'et\'any 1, Hungary. \email{gh@arago.elte.hu}}

\presentaddress{Department of Biological Physics, E\"otv\"os Lor\'and University, H-1117 Budapest, P\'azm\'any s\'et\'any 1, Hungary }

\abstract{Space debris larger than 1 cm can damage space instruments and impact Earth. The low-Earth orbits (at heights smaller than 2000 km) and orbits near the geostationary-Earth orbit (at 35786 km height) are especially endangered, because most satellites orbit at these latitudes. With current technology space debris smaller than 10 cm cannot be tracked. Smaller space debris burn up and evaporate in the atmosphere, but larger ones fall to the Earth's surface. For practical reasons it would be important to know the mass, composition, shape, velocity, direction of motion and impact time of space debris re-entering the atmosphere and falling to Earth. Since it is very difficult to measure these physical parameters, almost nothing is known about them. To partly fill this gap, we performed computer modelling with which we studied the celestial mechanics of spherical re-entry particles falling to Earth due to air drag. We determined the time, velocity and angle of impact as functions of the launch height, direction, speed and size of spherical re-entry particles. Our results can also be used for semi-spherical meteoroid particles of the interplanetary dust entering the Earth's atmosphere.}


\jnlcitation{\cname{%
\author{J. Sl\'{i}z-Balogh}, 
\author{D. Horv\'ath}, 
\author{R. Szab\'o}, and 
\author{G.  Horv\'ath}} (\cyear{2020}), 
\ctitle{Dynamics of spherical space debris of different sizes falling to Earth}, \cjournal{Astronomical Notes.}, \cvol{2020;00:1--6}.}


\maketitle


\section{Introduction}\label{sec1}

The number of satellites in the close orbits around Earth is increasing year by year. In 2017 economic, military and civil organizations launched more than 450 satellites - four times the yearly average number of satellites launched between 2000 and 2010~\citep{Witze2018}. These satellites over time will fail and become space debris, which at some time in the future will re-enter the atmosphere~\citep{Klinkrad2006a}. Between 1957 and 2002, more than 18000 trackable objects (66.7 \% of the total trackable objects orbiting the Earth) re-entered the atmosphere with a total mass of ~27000 tons (84 \% of the total mass of objects orbiting the Earth) and a total cross-sectional area of ~85000 $\mathrm{m}^2$ (67 \% of the total cross-sectional area of objects orbiting the Earth). 10-40 \% of the mass of larger objects may survive the severe structural and thermal loads during atmospheric descent to ground impact~\citep{Ailor2005}.
\par
The most common sources of space debris originate from spent fuel tanks, rocket cover plates, solar panel shards, fragments of deliberate and unintentional on-orbit explosions as well as the communal waste of space stations~\citep{Crowther2003}. In 2005 more than 14000 space objects with diameters larger than 5-10 cm could be tracked, of which only 5 \% were operational spacecraft~\citep{Klinkrad2006a}). In 2018 the number had increased to 23000. The larger space objects catalogued in 2018 had a mass of 6800 tons and 95 \% of it were dead satellites or pieces of inactive ones~\citep{Witze2018}.
\par
Both intentional and unintentional events can drastically increase the quantity of space debris when (i) a satellite explodes (e.g., on 11 January 2007 a Chinese satellite has been blown up in a missile test), or (ii) two space objects crash with each other (e.g., on 11 February 2009 a US commercial Iridium satellite smashed into the inactive Russian communications satellite Cosmos-2251)~\citep{Witze2018}. Even purposeful introduction of spherical objects has been performed/planned: (iii) In December 2006, the STS-116 Discovery space shuttle performed the ANDE (Atmospheric Neutral Density Experiment) mission. The ANDE satellite suite consisted of a series of four spherical micro-satellites with instrumentation to monitor the total neutral density along the orbit (300-400 km) for improved orbit determination of resident space objects, and to provide a test object for both radar and optical SSN (Space Surveillance Network) sensors~\citep{Nicholas2009}. (iv) The Japanese company Astro Live Experiences designed a system of small satellites scheduled for operation in 2020 for the purpose of creating an artificial, colorful meteor shower by releasing spherical pellets~\citep{Greshko2016}. The rise of debris from these mentioned events and others risks blocking space-based electronic communication. Each fresh impact increases exponentially the number possibilities for future impacts, a chain-reaction commonly known as the Kessler's syndrome~\citep{Kessler1978,Kessler1991}.
\par
Space debris with 1-10 cm size are especially dangerous, because they are impossible to track and their high kinetic energy is hard to shield against. Most satellites orbit at 2000 km above the Earth's surface in low-earth orbits, or at 35786 km altitude in geostationary orbits~\citep{Crowther2003}. The characteristic orbital speeds at these altitudes are 8 km/s and 3 km/s respectively. Due to this high speed, the resulting kinetic energy of the debris is hard to absorb even for a multilayered Whipple shield. The resulting problem is that space instruments can be damaged.
\par
The information about space debris and their behavior has increased progressively in the last two decades.~\cite{Klinkrad1993,Regan1993,Smirnov2001} developed space debris prediction and analysis models to determine the evolution of space debris from formation to destruction. They also modelled possible collisions of debris fragments of different sizes.~\citet{Crowther2003} reviewed the knowledge about space debris and proposed some short- and long-term potential ways to reduce their amount.~\citet{Klinkrad2006a} summarized and reviewed the scientific background of the dynamics and evolution of space debris and predicted the risks of debris to the environment.~\citet{Deleflie2008} analyzed the long-term (over 200 years) stability of space debris trajectories. They integrated both numerically and analytically the equations of motion of debris fragments including luni-solar gravitational perturbations.~\citet{Klinkrad2010} listed the sources and distribution of space debris fragments as well as their predicted evolution. He emphasized that one of the greatest problems with space debris is that they threaten the success of space missions. ~\citet{Leonov2011} studied the burning process of space debris in the Earth's atmosphere with the use of a meteor-monitoring television system. Using a symplectic integrator model,~\citet{Hubaux2012} simulated the motion trajectories and dynamics of space debris with high-Earth orbits including the geostationary ones. They concluded that space debris fragments can survive in their orbits for hundreds of years. In addition to the Earth's gravitational potential, their model considered luni-solar and planetary gravitational perturbations as well as solar radiation pressure.~\citet{Aslanov2014} modelled and analyzed the behavior of towed large space debris influenced by atmospheric disturbances of the Earth to reveal different methods to reduce the amount of space debris.~\citet{Aslanov2015} reported about active debris-removal systems. They showed that the flexible appendages of towed space debris influence the debris system, but without high amplitude vibrations. Using a symplectic orbital propagator,~\citet{DiCintio2017} modelled the long-term evolution of space debris ranging from low- to high-Earth orbits. Their simulation took also into consideration luni-solar perturbations.
\par
Because of the threat to working satellites space debris larger than 10 cm is tracked with ground-based radar and optical networks that observe, catalogue and monitor the positions and orbits of these objects~\citep{Stansbery1997,Goldstein1998,Klinkrad2010,Witze2018}. If it is necessary, the endangered satellites can perform space-debris-avoiding maneuvers. However, space debris smaller than 10 cm remains hidden to low-Earth orbit tracking systems.
\par
Depending on several parameters, different re-entry objects orbit around Earth for various time periods (called impact time further on in this work) and hit the Earth with different speeds. For practical reasons it would be important to know the mass, composition, shape, velocity, direction of motion and impact time of re-entry objects falling on Earth. Since it is very difficult to measure these physical parameters, almost nothing is known about them. To partly fill this gap, we performed computer modelling with which we studied the celestial mechanics of spherical re-entry objects falling on Earth due to air drag. We determined the time, velocity and angle of impact as functions of the launch height, direction, speed and size of spherical re-entry particles. Our results can also be used for semi-spherical meteoroid particles of the interplanetary dust~\citep{Grun2001} entering the Earth's atmosphere. Although numerous different analytical techniques and computational methods have been developed to investigate the motion, trajectories and evolution of satellites and space debris with or without atmospheric drag~\citep{Cook1972,Aksnes1976,Liu1979,King-Hele1987,Brumberg1995, Rossi1995,Montenbruck2000,Beutler2005a,Beutler2005b,Klinkrad2006b,Salnikova2018,Salnikova2019} the computational studies presented here have not been performed yet.

\section{Model and Methods}\label{sec2}

\subsection{Calculation of air density versus height}

\begin{figure}[t]
\includegraphics[width=80mm]{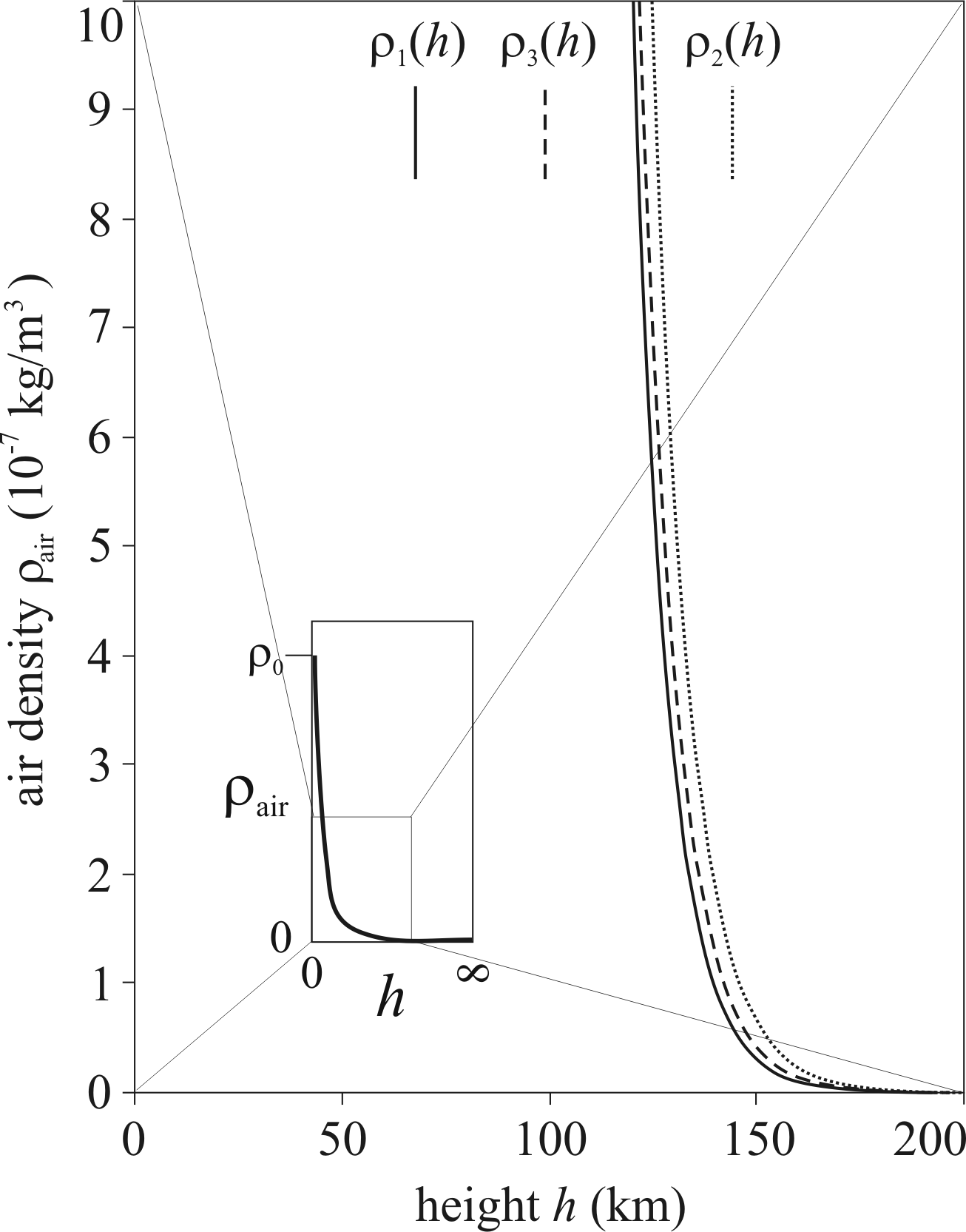}
\caption{Air density $\rho_{air}(\textit{h}$) versus height \textit{h} above the Earth's surface calculated with the use of the formulae $\rho_1(\textit{h}$), $\rho_2(\textit{h}$ and $\rho_3(\textit{h}$) described by equation~(\ref{eq:ro1}),~(\ref{eq:ro2}) and~(\ref{eq:ro3}), respectively. In the inset the qualitative function $\rho_{air}(\textit{h}$) is represented, an enlarged part of which is shown by the large figure.}\label{fig1}
\end{figure}

Depending on the spatial change of the gravitational acceleration and the geometry of the static, exponential atmosphere, the air density  $\rho_{air}$ as a function of the height \textit{h} above the Earth's surface is the following~\citep{Klinkrad2006c}:

for constant gravitational acceleration in planar atmosphere:

\begin{eqnarray}
\rho_{1}(h) = \rho_{0} e^{\frac{-M\gamma {m_{E}}}{RT{r^{2}_{E}}}h}
\label{eq:ro1}
\end{eqnarray}

for changing gravitational acceleration in planar atmosphere:

\begin{eqnarray}
\rho_{2}(h) = \rho_{0} e^{\frac{-M\gamma {m_{E}}}{RT}\left(\frac{1}{r_E}- \frac{1}{r_E+h}   \right) }
\label{eq:ro2}
\end{eqnarray}

for changing gravitational acceleration in spherical atmosphere:

\begin{eqnarray}
\rho_{3}(h) = \rho_{0}\left(\frac{r_E}{r_E+h}\right)^2 e^{\frac{-M\gamma {m_{E}}}{RT}\left(\frac{1}{r_E}- \frac{1}{r_E+h}   \right) }
\label{eq:ro3}
\end{eqnarray}

where $\rho_0 = 1.23  \mathrm{kg/m}^3$ is the air density on the Earth's surface under normal conditions (temperature \textit{T} = 300 K, press \textit{p} = 1 bar), $ r_E = 6.371\cdot10^6$ m is the Earth's average radius, $\textit{M} = 29\cdot10^{-3}$ kg is the molar mass of air, $\gamma = 6.67408\cdot10^{-11}  \mathrm{m^3kg^{-1}s^{-2}}$ is the universal gravitational constant, $\textit{m}_E = 5.972\cdot10^{24}$ kg is the Earth's mass, \textit{R} = 8.314 J/K/mol is the universal gas constant. Fig.~\ref{fig1} shows the air density functions  $\rho_{air}$(\textit{h}) calculated with the use of ~(\ref{eq:ro1}), ~(\ref{eq:ro2}) and ~(\ref{eq:ro3}). Since there are only negligible differences among the numerical values of these three functions at a given \textit{h}-value, in our computer model we used the simplest formula ~(\ref{eq:ro1}) for the height-dependent air density. For this reason, it would not have been worth using a more precise (non-static and non-exponential) atmosphere model, such as the U. S. Standard Atmosphere 1976, for example.

\begin{figure}[t]
\includegraphics[width=80mm]{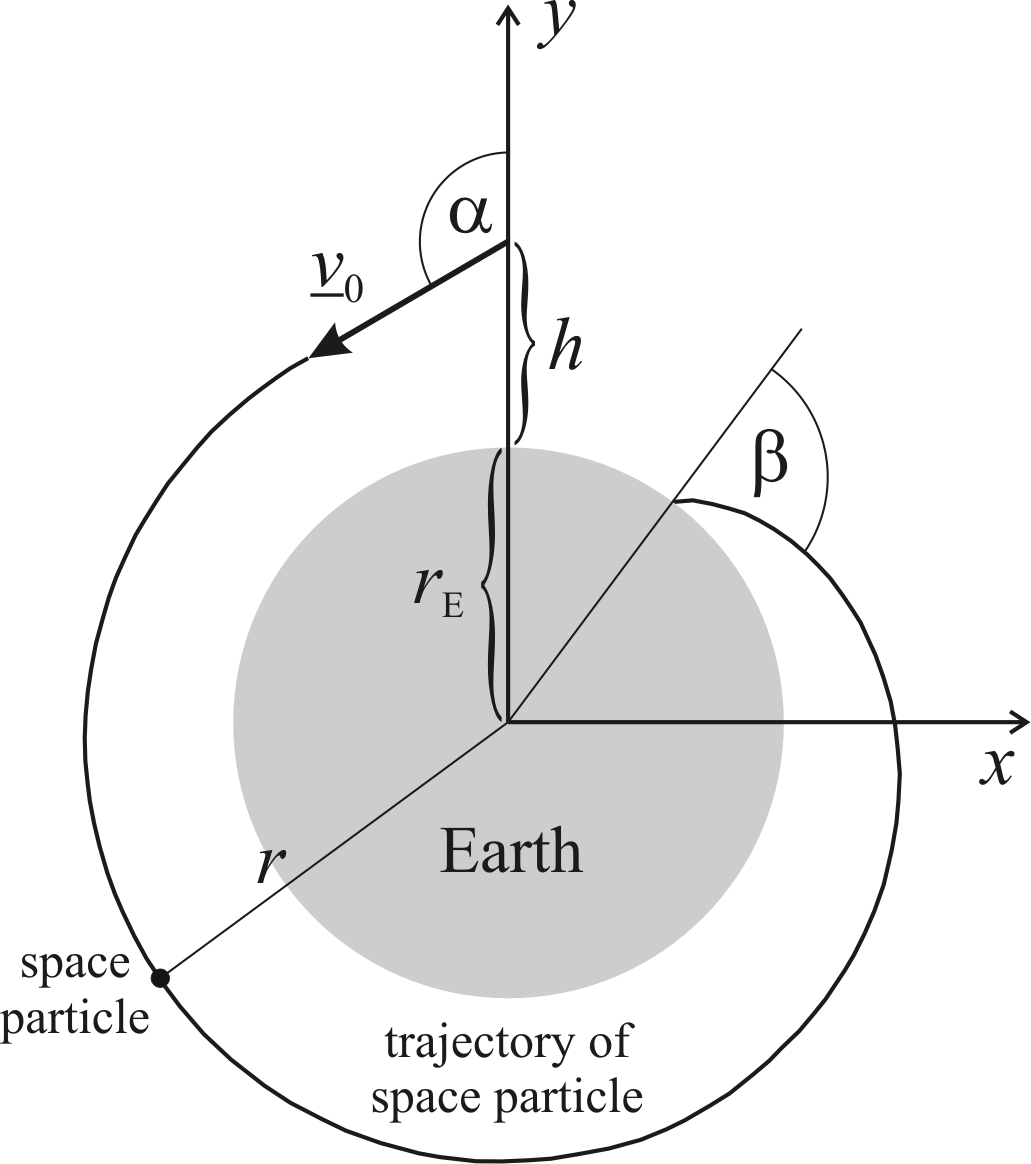}
\caption{Geometry of a two-dimensional ($\textit{x, y}$) computer model, in which a spherical space particle is launched with an initial velocity vector $\vec{v}_0$ and angle $\alpha$ from the radial direction at height $\textit{h}$ above the Earth's surface and falls to the Earth's surface with impact angle $\beta$.$\textit{r}_E$: radius of the Earth, \textit{r}: distance of the particle from the Earth's center.}\label{fig2}
\end{figure}

\begin{figure*}[t]
\centerline{\includegraphics[width=160mm]{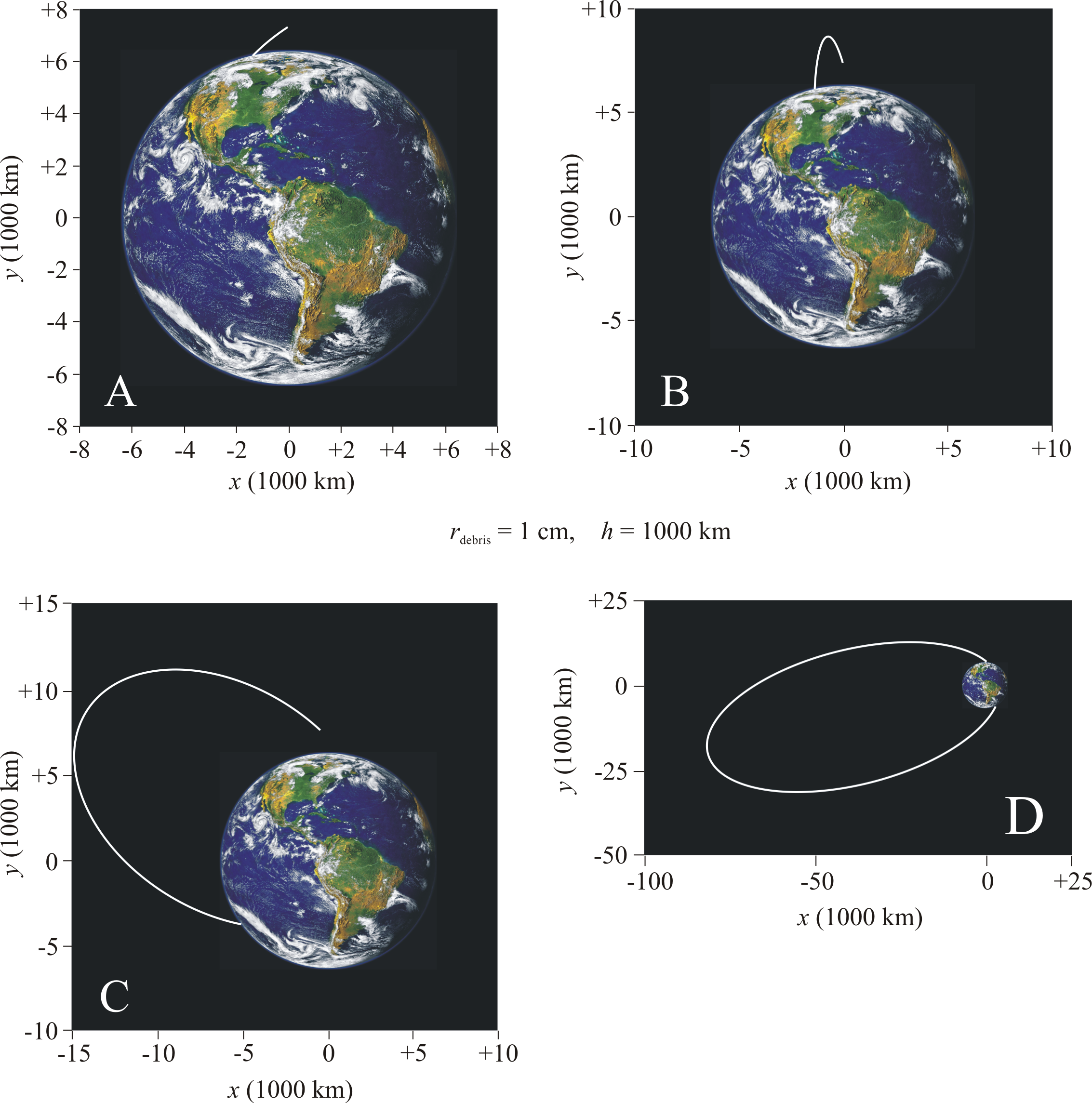}}
\caption{Some typical trajectories of spherical iron particles $\left(7.9\cdot10^3 \mathrm{kg}/\mathrm{m}^3\right)$ with radius $r_{debris}$ = 1 cm launched from height $h = 1000$ km above the Earth's surface with different initial velocities $v_0$ and angles $\alpha$ from the radial direction: (A) $v_0 = 6.87$ km/s, $\alpha = 122.6^\circ$, (B)$ v_0 = 4.1$ km/s, $\alpha = 16.38^\circ$, (C) $v_0 = 8.18$ km/s, $\alpha = 45^\circ$, (D) $v_0 = 9.95$ km/s, $ \alpha = 54.4^\circ$. The time for the particles to fall to the Earth's surface is 7.2 min (A), 27.4 min (B), 135.4 min (C), 1540.8 min (D).\label{fig3}}
\end{figure*}

\begin{figure*}[t]
\centerline{\includegraphics[width=160mm]{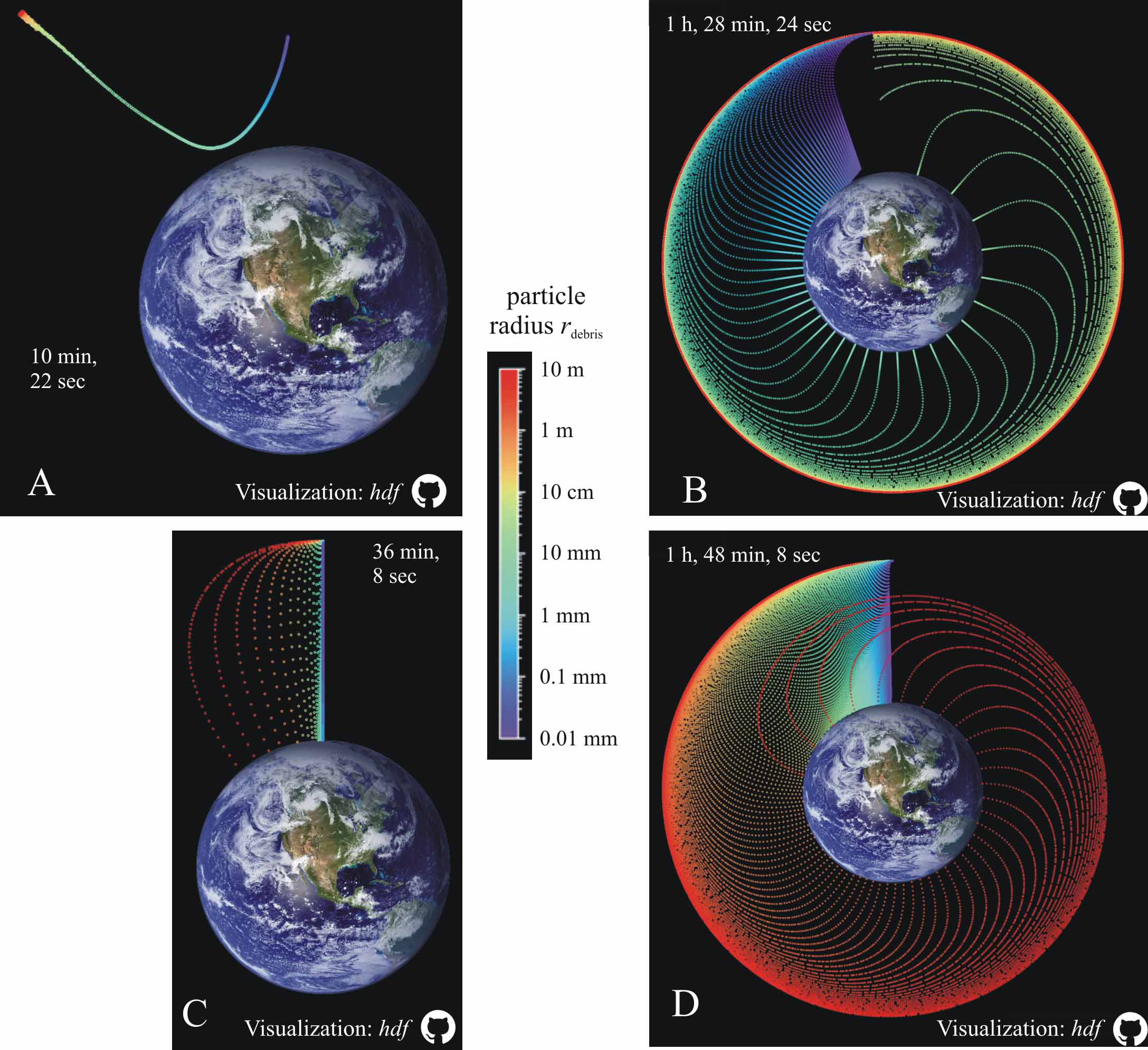}}
\caption{(A) Positions (colored dots) of 1000 spherical iron particles $\left(7.9\cdot10^3 \mathrm{kg}/\mathrm{m}^3\right)$ with different radius $r_{debris}$ at 10 min 22 sec after their launch with tangential $\left(\alpha = 90^\circ\right)$ circular orbit velocity $v_0 = 7.847$ km/s at the same point from height $h = 100$ km from the Earth's surface (see Supplementary Video Clip VC1). The region above the Earth's surface $\left(h \ge 0 \right)$ is for illustrative purposes magnified 100 times. (B) Trajectories (series of dots of a given color) of 100 particles with different radius $r_{debris}$ at 1 h 28 min 24 sec after their launch with tangential $\left(\alpha = 90^\circ\right)$  circular orbit velocity $v_0 = 7.817$ km/s from height $h = 150$ km (see Supplementary Video Clip VC2). The region above the Earth's surface is 66 times magnified. (C) Trajectories of 20 particles with different radius $r_{debris}$ at 36 min 8 sec after their launch with tangential $\left(\alpha = 90^\circ\right)$ circular orbit velocity $v_0 = 7.877$ km/s from height $h = 50$ km (see Supplementary Video Clip VC3). The region above the Earth's surface is magnified 200 times. (D) Trajectories of 100 particles with different radius $r_{debris}$  at 1 h 48 min 8 sec after their launch with tangential $\left(\alpha = 90^\circ\right)$ circular orbit velocity $v_0 = 7.847$ km/s from height $h = 100$ km (see Supplementary Video Clip VC4). The region above the Earth's surface is magnified 100 times. Different radii $r_{debris}$ are coded with different colors from violet for the smallest $\left(0.01 \mathrm{mm}\right)$ to red for the largest $\left(10^4  \mathrm{mm}\right)$ particle.\label{fig4}}
\end{figure*}

\begin{figure*}[t]
\centerline{\includegraphics[width=160mm]{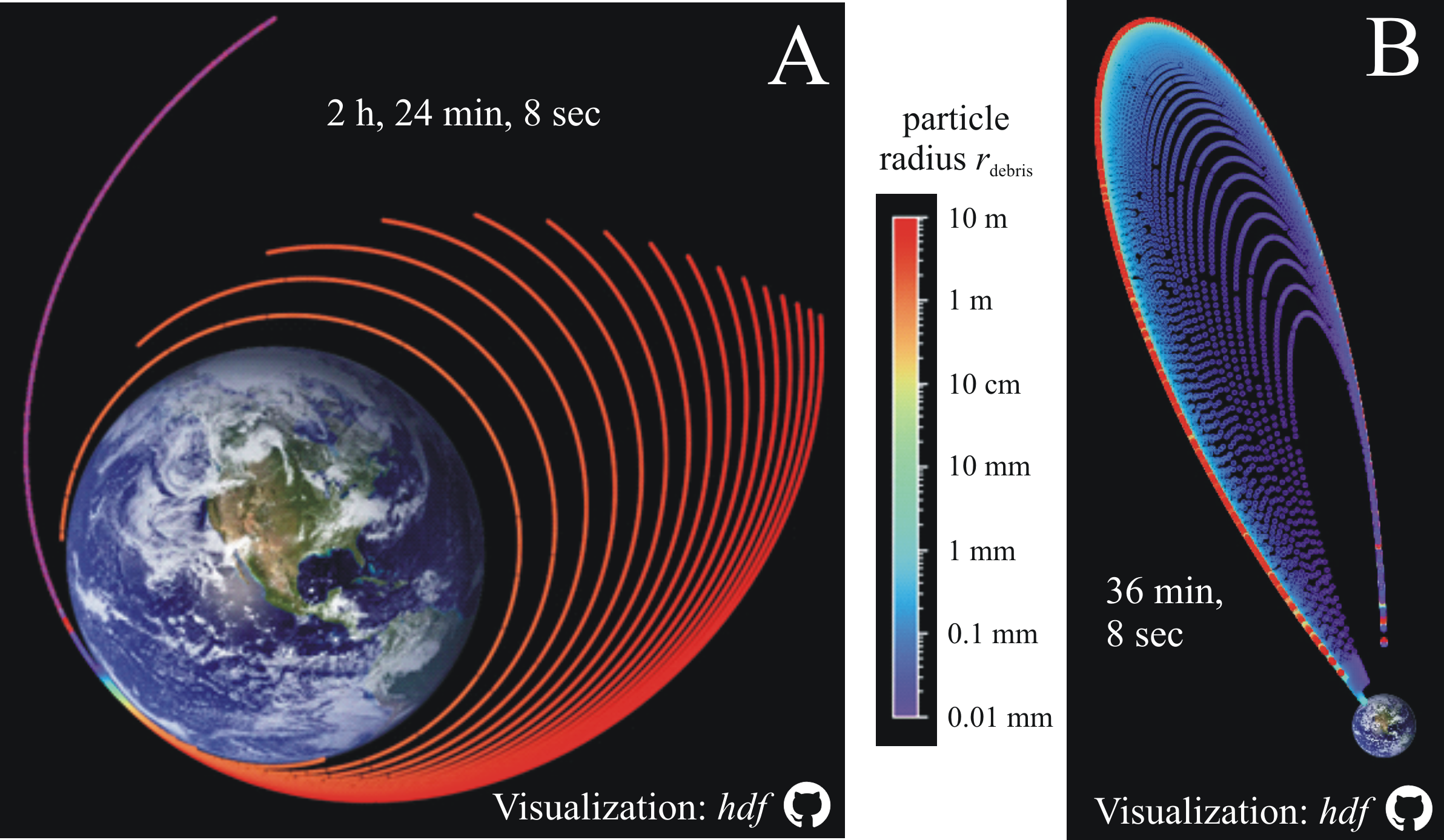}}
\caption{(A) Trajectories of 100 spherical iron particles $\left(7.9\cdot10^3 \mathrm{kg}/\mathrm{m}^3\right)$ with different radius $r_{debris}$ at 2 h 24 min 8 sec after their launch with an initial velocity $v_0 = 4.57$ km/s and angle $\alpha = 122.6^\circ$ relative to the radial direction from the same point at height $h = 10000$ km above the Earth's surface (see Supplementary Video Clip VC5). The region above the Earth's surface $\left(h \ge 0 \right)$ is shown to scale. (B) Trajectories of 100 particles with different radius $r_{debris}$ at 36 min 8 sec after their launch with initial velocity $v_0 = 5.65$ km/s and angle $\alpha = 45^\circ$ from height $h = 100$ km (see Supplementary Video Clip VC6). The region above the Earth's surface is magnified 100 times for illustrative purposes. Different radii $r_{debris}$ are coded with different colors from violet for the smallest (0.01 mm) to red for the largest ($10^4$ mm) particle.\label{fig5}}
\end{figure*}

\begin{figure}[t]
\includegraphics[width=80mm]{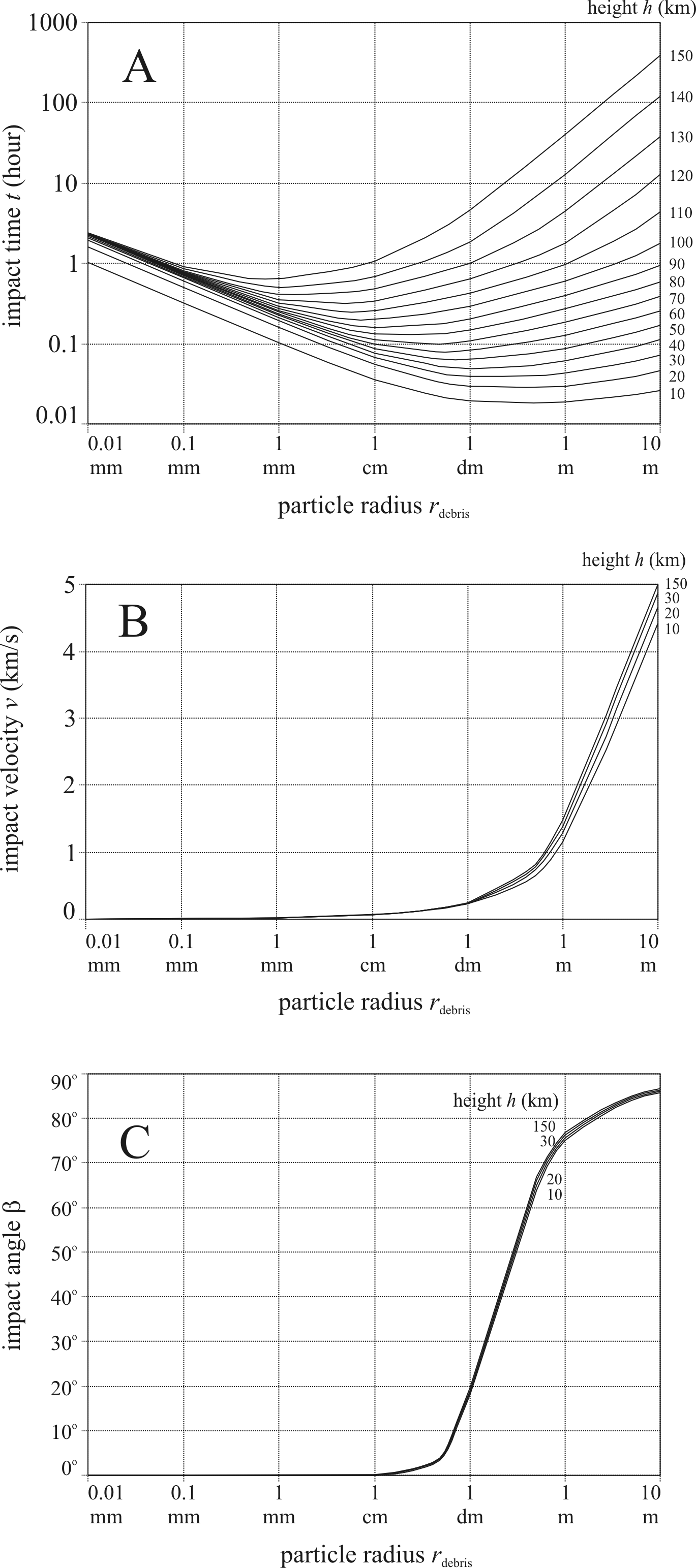}
\caption{(A) Impact time $t$ as a function of the radius $r_{debris}$ of iron particles $\left(7.9\cdot10^3 \mathrm{kg}/\mathrm{m}^3\right)$ launched with tangential ($\alpha = 90^\circ$) each with its circular orbit velocity $v_0$ (see Table\ref{tab1}) from different heights h from the Earth's surface. (B) Impact velocity $v$ as a function of the radius $r_{debris}$ of particles launched with tangential ($\alpha = 90^\circ$) each with its circular orbit velocity $v_0$ (see Table.~\ref{tab1}) from different launching heights $h$. (C) Impact angle $\beta$ from the local vertical (perpendicular to the Earth's surface) as a function of the radius $r_{debris}$ of particles launched with tangential ($\alpha = 90^\circ$) each with its circular orbit velocity $v_0$ (see Table\ref{tab1}) from different launching heights $h$.}\label{fig6}
\end{figure}

\begin{figure*}[t]
\centerline{\includegraphics[width=160mm]{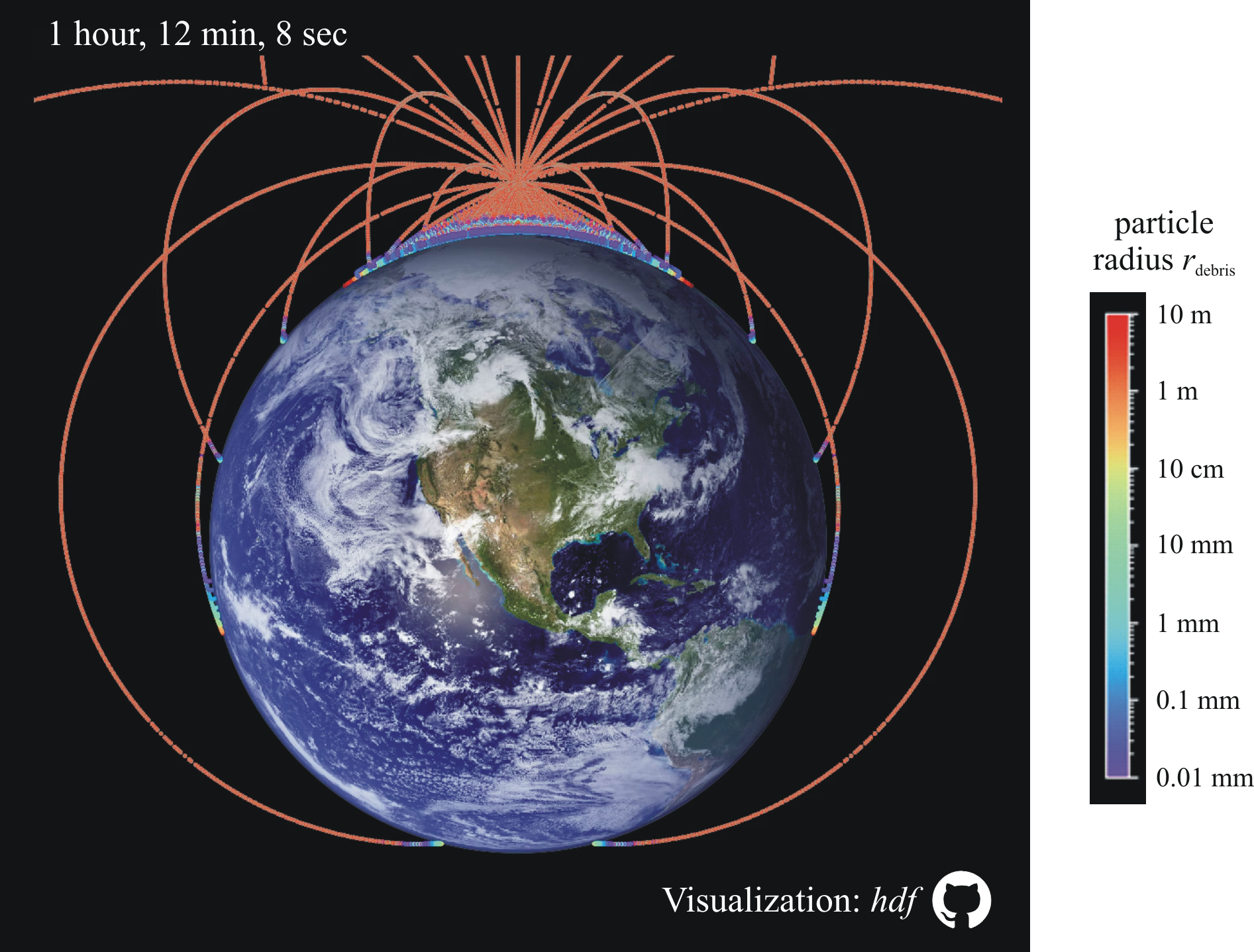}}
\caption{A simulation of an explosion event. Trajectories of 30 (different angles $\alpha) \times$ 11 (different radii $r$) = 330 spherical iron particles $\left(7.9\cdot10^3 \mathrm{kg}/\mathrm{m}^3\right)$ at 1 h 12 min 8 sec after the explosion at height $h = 1000$ km above the Earth's surface (see Supplementary Video Clip VC7). The initial velocity of all explosion fragments was $v_0 = 7$ km/s and the angle of the initial velocity vector changed from $\alpha = 0^\circ$ to $\alpha = 360^\circ$ with an increment $\Delta \alpha = 12^\circ$ (see Supplementary Video Clip VC7). The region above the Earth's surface$\left(h \ge 0 \right)$ is not magnified in the visualization. Different particle radii $r_{debris}$ are coded with different colors from violet for the smallest (0.01 mm) to red for the largest ($10^4$ mm) particle.\label{fig7}}
\end{figure*}

\begin{figure*}[t]
\centerline{\includegraphics[width=160mm]{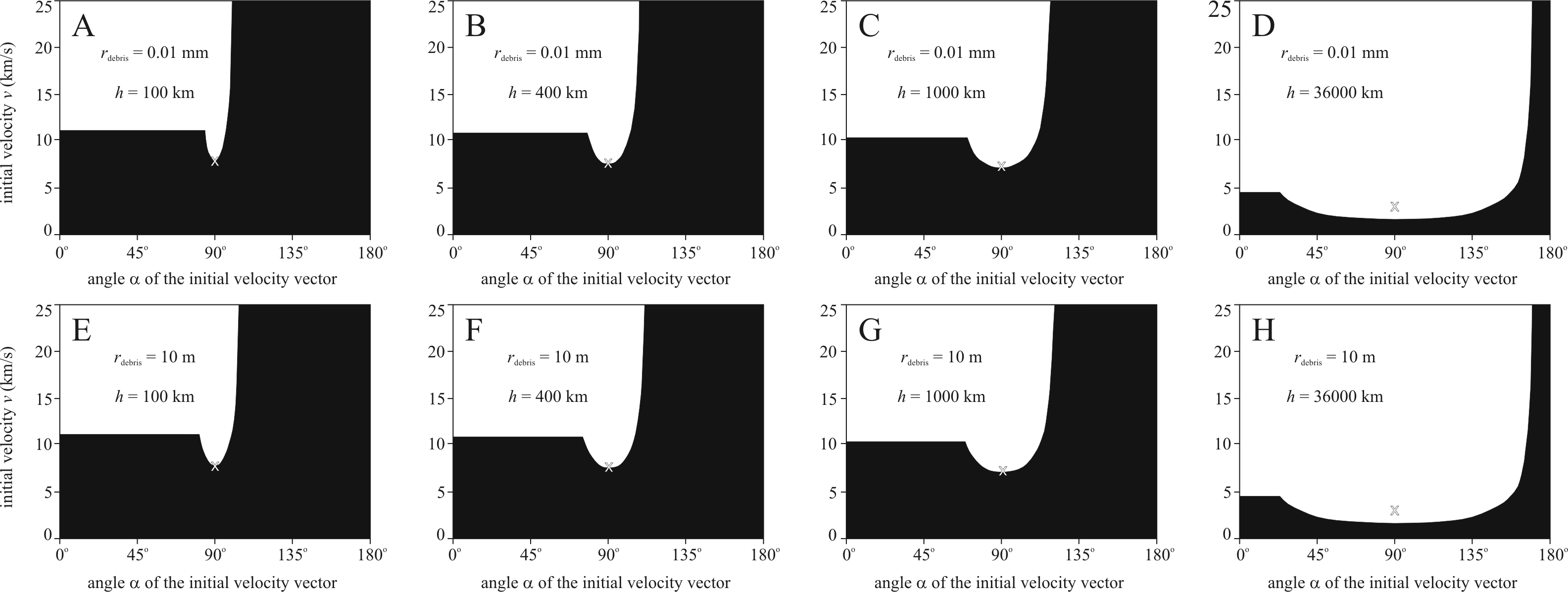}}
\caption{Simulation of an explosion event. At height $h = 100, 400, 1000$ and 36000 km above the Earth's surface, spherical iron particles$\left(7.9\cdot10^3 \mathrm{kg}/\mathrm{m}^3\right)$ of radii $r_{debris}$ = 0.01 mm and 10 m start with velocities 0 km/s $\leq v_0 \leq 23$ km/s and angles $\alpha = 0^\circ$ to $\alpha = 180^\circ$ (from the local radial direction). If the initial velocity vector $\vec{v}_0(v_0,\alpha)$ falls on the black region of the $ v_0-\alpha$ map, then the explosion fragments fall to the Earth's surface within 50 days. Symbol x denotes the tangential ($\alpha = 90^\circ$) circular orbit velocity at a given height $h$.\label{fig8}}
\end{figure*}

\subsection{Equation of motion of spherical particles in the Earth's atmosphere}

In our computer model, every fragment of space debris was a sphere with radius $r_{debris}$ and constant density $\rho_{debris} = 7.9\cdot10^3 \mathrm{kg/m^3}$ of iron. We modelled only spherical particles, because this shape has the simplest aerodynamics. Considering other shapes would result in a much more complicated, shape-dependent, temporally changing air drag force vector described by the Navier-Stocks equations.
\par
We did not model the burning of space particles in the atmosphere. We assumed that the size (radius) of a given orbiting and falling particle is constant. The burning is a very complex thermo- and aerodynamical chemical process depending on the shape and composition of the particle as well as on the local composition and density of the atmosphere~\citep{Allen1953,Bouslog1994,Fritsche1999,Fritsche2000,Koppenwallner2001,Klinkrad2004,Rochelle2004,Klinkrad2006}. Involving the particle burning into our model could be an important task of a separate paper.
\par
We simulated the motion of particles falling from heights \textit{h} = 50, 100, 150, 400, 1000, 10000 and 36000 km above the Earth's surface. These are reasonable heights for the origins of space debris, since in the low Earth orbits manned missions are mostly below 400 km whereas Earth observation satellites operate between 800 and 1500 km and above that region is the geostationary orbit (\textit{h} = 35786 km).
\par
As one of the possible groups of direction (angle $\alpha$) of the initial velocity vector $\vec{v}_0$ of space particles (Fig.~\ref{fig2}) was assumed to be nearly tangential ($\alpha \approx 90^\circ$). This case modelled the situation when two space objects moving in two different orbits cross each other and the faster object collides with the slower one. Such a space event results in numerous fragments with $\alpha \approx 90^\circ$ initial angle.
\par
As another typical possibility of the direction of initial velocity vector $\vec{v}_0$ of space debris particles was assumed that $\alpha$ changes from $0^\circ$ to $360^\circ$ with an increment $\Delta\alpha = 12^\circ$. This modelled either the situation after a crash of two space objects moving in the same orbit with opposite velocity vectors, or the explosion of a satellite. Both space events result in numerous fragments flying away in all possible directions.
\par
In the two-dimensional ($\textit{x, y}$) coordinate system of  Fig.~\ref{fig2}, a spherical space particle was launched with an initial velocity vector $\vec{v}$ and angle $\alpha$ from the radial direction at height  $\textit{h}$. The potential energy of this particle is

\begin{eqnarray}
U_{debris} = -\frac{\gamma m_{debris} m_E}{r} ,    
\label{eq:udebris}   
\end{eqnarray}

where $\textit{r}$ is the distance of the particle from the Earth's center and

\begin{eqnarray}
m_{debris} = \frac{4\pi \rho_{debris}r^3_{debris}}{3} ,      
\label{eq:mdebris}    
\end{eqnarray}

is the mass of the particle. The magnitude of the air drag is

\begin{eqnarray}
S =\frac{Ac\rho_{air}v^2}{2} ,    
\label{eq:s}      
\end{eqnarray}

where $\rho_{air}$ is the air density, $\textit{v}$ is the velocity of the spherical particle, $\textit{c}$ = 0.4 is the drag coefficient (or shape factor) of the sphere and

\begin{eqnarray}
A =\pi r^2_{debris} ,  
\label{eq:A}         
\end{eqnarray}

is the effective cross-sectional area of the sphere. The equation of motion for the $\vec{ r}(t) = \left[x(t), y(t)\right]$ site vector of the particle in the gravitational field and atmosphere of the Earth are the following:

\begin{eqnarray}
m_{debris}\frac{\mathrm{d}^2\vec{r}}{\mathrm{d}t^2} =\vec{G}+\vec{S} ,  
\label{eq:mdebris2}         
\end{eqnarray}

where

\begin{eqnarray}
\vec{G} =-\frac{\gamma m_{debris}m_E}{r^2}\frac{\vec{r}}{r}   
\label{eq:g}         
\end{eqnarray}

is the gravitational force and

\begin{eqnarray}
\vec{S} =-\frac{0.4\pi r^2_{debris}v^2}{2}\rho_0 e^{-\frac{M\gamma m_E}{RTr^2_E}h}\frac{\vec{v}}{v} ,  
\label{eq:ss}         
\end{eqnarray}

is the drag force as a function of height

\begin{eqnarray}
h=r-r_E ,  
\label{eq:h}         
\end{eqnarray}

above the Earth's surface. Using ~(\ref{eq:mdebris2}), ~(\ref{eq:g}), ~(\ref{eq:ss}) and ~(\ref{eq:h}), the equation of motion becomes:

\begin{eqnarray}
\frac{\mathrm{d}^2\vec{r}}{\mathrm{d}t^2} =-\frac{\gamma m_E}{r^3}\vec{r}-\frac{0.15\rho_0}{\rho_{debris}r_{debris}}e^{-\frac{M\gamma m_E\left(r-r_E\right)}{RTr^2_E}}v\vec{v} .  
\label{eq:mozg}         
\end{eqnarray}

Using ~(\ref{eq:mozg})), the equation of motion for the \textit{x(t}) and \textit{y(t)} coordinates of the particle in the system of coordinates of  Fig.~\ref{fig2} are the following:

\begin{eqnarray}
\frac{\mathrm{d}^2 x\left(t\right)}{\mathrm{d}t^2} =-\frac{\gamma m_E}{r(t)^3}x\left(t\right)-\frac{0.15\rho_0}{\rho_{debris}r_{debris}}e^{-\frac{M\gamma m_E\left[r(t)-r_E\right]}{RTr^2_E}}v\left(t\right)v_x\left(t\right) ,  
\label{eq:mozgx}         
\end{eqnarray}

\begin{eqnarray}
\frac{\mathrm{d}^2 y\left(t\right)}{\mathrm{d}t^2} =-\frac{\gamma m_E}{r(t)^3}y\left(t\right)-\frac{0.15\rho_0}{\rho_{debris}r_{debris}}e^{-\frac{M\gamma m_E\left[r(t)-r_E\right]}{RTr^2_E}}v\left(t\right)v_y\left(t\right),  
\label{eq:mozgy}         
\end{eqnarray}

\begin{eqnarray}
r\left(t\right) =\sqrt{{x\left(t\right)}^2 +{y\left(t\right)}^2},  v\left(t\right) =\sqrt{{v_x\left(t\right)}^2 +{v_y\left(t\right)}^2},  
\label{eq:rt}         
\end{eqnarray}

\begin{eqnarray}
v_x\left(t\right)= \frac{\mathrm{d}x\left(t\right)}{\mathrm{d}t} , v_y\left(t\right)= \frac{\mathrm{d}y\left(t\right)}{\mathrm{d}t}
\label{eq:v}         
\end{eqnarray}

with $\gamma = 6.67408\cdot10^{-11}\mathrm{m}^3\mathrm{kg}^{-1}\mathrm{s}^{-2}$, $m_E = 5.972\cdot10^{24}$ kg, $r_E = 6.371\cdot10^6$ m, $\rho_0 = 1.23\mathrm{ kg}/\mathrm{m}^3$,$\rho_{debris} = 7.9\cdot10^3 \mathrm{kg}/\mathrm{m}^3$, $M = 29\cdot10^{-3}$ kg, $R = 8.314$ J/K/mol, $T = 300$ K. The equation of motion ~(\ref{eq:v}) were solved numerically with the use of Runge-Kutta-Fehlberg 7(8) integrator ~\citep{Fehlberg1968} in which the actual step size was determined according to the desired accuracy $\epsilon = 10^{-16}$ (= tolerated local error per unit step). The accuracy of the Runge-Kutta-Fehlberg 7(8) method including stepsize control was sufficiently accurate for this task. First we performed our calculations with Runge-Kutta 4(5) then with Runge-Kutta-Fehberg 7(8) method. The coordinates obtained by these two methods differed only after the 6th decimal.
\par
In a preliminary simulation, we took into consideration the non-spherical geoid shape of the Earth (~\citet{Klinkrad2006d}, see also Electronic Supplementary Material), but found the geoid had only a negligible (0.01 \%) influence on the impact time of space particles. Thus, we found it reasonable to simplify computations by approximating Earth as spherical.

\begin{center}
\begin{table*}[t]%
\centering
\setlength{\tabcolsep}{4pt}
\caption{The circular orbit velocity $v_0(h)= \sqrt{\gamma m_E/(r_E+h)}$ (km/s) as a function of the height $h$ (km) above the Earth's surface.}
\begin{tabular}{lllllllllllllllll}\hline
$v_0$ (km/s) & 7.902&7.896&7.890&7.883&7.877&7.871&7.865&7.859&7.853&7.847&7.841&7.835&7.829&7.823&7.817\\
$h$ (km) & 10 & 20 & 30 & 40 & 50 & 60 & 70 & 80 & 90 & 100 & 110 & 120 & 130 & 140 & 150\\\hline
\end{tabular}
\label{tab1}
\end{table*}
\end{center}

\begin{center}
\begin{table*}[t]%
\caption{Mean impact time <t> (minutes) of spherical iron particles with radius $r_{debris}$ and density of $\left(7.9\cdot10^3 \mathrm{kg}/\mathrm{m}^3\right)$ of a space debris exploded at height $h$ above the Earth's surface. The components $v_0$ and $\alpha$ (angle measured from the local radial direction) of the initial velocity vector $\vec{v}_0(v_0,\alpha)$ of the particles changed from $v_0 = 0$ to 23 km/s with an increment $ \Delta v_0 = 0.23$ km/s and $\alpha = 0^\circ$ to $360^\circ$ with an increment $\Delta\alpha = 3.6^\circ$.\label{tab2}}
\centering
\begin{tabular*}{500pt}{@{\extracolsep\fill}lccD{.}{.}{3}c@{\extracolsep\fill}}
\toprule
&\multicolumn{3}{@{}c@{}}{\textbf{H}} \\\cmidrule{2-4}
\textbf{$r_{debris}$} & \textbf{100 km}  & \textbf{400 km}  & \textbf{1000 km}  & \textbf{36000 km}   \\
\midrule
\textbf{0.01 mm} & 316.8 min&	313.3 min & 314.4  \hspace*{0.3333em}\mathrm{min} & 545.6 min   \\
\textbf{10 m} & 134.5 min	&156.5 min & 143.0  \hspace*{0.3333em} \mathrm{min} &401.2 min \\
\bottomrule
\end{tabular*}
\end{table*}
\end{center}

\section{Results}\label{sec3}

 Fig.~\ref{fig3} shows some typical trajectories of spherical space particles with radius $r_{debris}$ = 1 cm launched from a height $h$ = 1000 km above the Earth's surface with different initial velocities $v_0$ and angles $\alpha$. Outside the Earth's atmosphere (practically higher than 300 km) these trajectories are elliptical (because $v_0 < 10.446$ km/s = escape speed at height $ h$ = 1000 km), but they become ballistic after the space particle entered denser air layers. After launched, the particle falls to the Earth's surface in time $t$ called impact time. This impact time is smaller (7.2, 27.4 min) or larger (135.4, 1540.8 min), if the trajectory of the particle is shorter ( Fig.~\ref{fig3}A,B) or longer (Fig.~\ref{fig3}C,D), respectively.
\par
Fig.~\ref{fig4}A shows the positions of 1000 spherical space particles at 10 min 22 sec after their launch. The color-coding of the radius $r_{debris}$ helps to visualize how the particles of various sizes displace over this time. The particles are launched with a tangential ($\alpha = 90^\circ$) velocity $v_0$ = 7.847 km/s from the same point at height $h$ = 100 km. For illustration purposes the region above the Earth is not to scale, it has been magnified 100 times, but not magnified in the calculations. Although all the 1000 particles started from the same point, their positions became different due to the size-dependent air drag. Because of this size dispersion of the trajectory, the series of the different positions of space particles formed a convex semi-parabolic chain (Fig.~\ref{fig4}A). The lowermost part of this chain is greenish, which means particle radii 5 mm < $r_{debris}$ < 50 mm. Therefore, these particles fall first to the Earth's surface (see Supplementary Video Clip VC1).
\par
Fig.~\ref{fig4}B displays the trajectories of 100 space particles with different color-coded radius $r_{debris}$ at 1 h 28 min 24 sec after their launch with tangential ($\alpha = 90^\circ$) velocity $v_0$ = 7.817 km/s from height $h$ = 150 km. The region above the Earth's surface has been magnified 66 times. The different trajectories of the different-sized particles induced by the size-dependent air drag are clearly seen. All trajectories seem to touch the Earth's surface nearly perpendicularly (impact angle $\beta \approx 0^\circ$). However, such small impact angles are only a visualization artefact induced by the 66-200 times enlargement of the atmosphere thickness. Angle $\beta$ depends strongly on the particle radius $r_{debris}$, but is practically independent of the launch height $h$. First the particles with radii 5 mm < $r_{debris}$ < 50 mm (coded with green) fell to Earth, then smaller and smaller (coded with bluish-violet) as well as larger and larger (coded with orange-reddish) particles hit the surface, while finally the largest particles with $r_{debris}$ = 10 m (coded with red) collided with the Earth (see Supplementary Video Clip VC2).
\par
Fig.~\ref{fig4}C represents the trajectories of 20 space particles with different color-coded radius $r_{debris}$ at 36 min 8 sec after their launch with tangential ($\alpha = 90^\circ$) velocity $v_0$ = 7.877 km/s from height $h$ = 50 km. The region above the Earth's surface ($h \geq 0$) has been magnified 200 times for illustrative purposes. First the particles with radii 50 mm < $r_{debris}$ < 100 mm (coded with green and yellow) fell to Earth, then larger and larger (coded with orange and red) as well as smaller and smaller (coded with violet, blue) particles hit the surface, finally the smallest ($r_{debris}$ = 0.01 mm, coded with violet) particles impact the Earth (see Supplementary Video Clip VC3).
\par
Fig.~\ref{fig4}D shows the trajectories of 100 space particles with different color-coded radius $r_{debris}$ at 1 h 48 min 8 sec after their launch with tangential ($\alpha = 90^\circ$) velocity $v_0$ = 7.847 km/s from height $h$ = 100 km. The region above the Earth has been magnified 100 times. First the particles with radii 5 mm < $r_{debris}$ < 50 mm (coded with green) fell to Earth, then smaller and smaller (coded with green, blue, violet) as well as larger and larger (coded with yellow, orange, red) particles hit the surface, finally the largest ($r_{debris}$ = 10 m, coded with red) and smallest ($r_{debris}$ = 0.01 m, coded with violet) particles simultaneously impact the Earth (see Supplementary Video Clip VC4).
\par
Fig.~\ref{fig5}A represents the trajectories of 100 space particles with different color-coded radius $r_{debris}$ at 2 h 24 min 8 sec after their launch with initial velocity $v_0$ = 4.57 km/s and angle $\alpha = 122.6^\circ$ from the same point at height $h$ = 10000 km. The region above the Earth's surface is to scale. The trajectories are ellipses. The radial direction of the farthest points (apogee) of these ellipses from the Earth does not rotate. Until they enter the atmosphere, all particles with different sizes move together. Reaching the atmosphere, first the particles with radii 1 mm < $r_{debris}$ < 800 mm (coded with green and yellow) fell to Earth, then smaller and smaller (coded with blue, violet) particles hit the surface, while finally the largest particle with $r_{debris}$ = 10 m (coded with red) impacted Earth (see Supplementary Video Clip VC5).
\par
Fig.~\ref{fig5}B displays the trajectories of 100 space particles with different radius $r_{debris}$ at 36 min 8 sec after their launch with initial velocity $v_0$ = 5.65 km/s and angle $\alpha = 45^\circ$ from height $h$ = 100 km. The region above the Earth's surface has been magnified 100 times for better visualization of the events. Due to the initial angle $\alpha = 45^\circ$, the trajectories are elongated ellipses touching the Earth's surface after the particles entered the atmosphere. Similar to Fig.~\ref{fig4}C, here the particles with radii 1 m <$r_{debris}$ < 10 m (coded with red) first fall to the Earth's surface, then smaller and smaller (coded with blue, violet) particles hit the surface, while finally the smallest particle with $r_{debris}$ = 0.01 mm (coded with violet) collided with the Earth (see Supplementary Video Clip VC6).
\par
Fig.~\ref{fig6}A shows the impact time $t$ as a function of the radius $r_{debris}$ of space particles launched tangentially with respect to their orbit at a speed characteristic of the height of their orbit ($\alpha = 90^\circ$, $v_0(h)= \sqrt{\gamma m_E/(r_E+h)}$ (km/s) , see Table~\ref{tab1}). According to Fig.~\ref{fig6}A, the radius $r^*_{debris}$ of the particle that will first fall to the Earth increases from 1 to 500 mm as the launching height decreases from 150 to 10 km. A gradual increase or decrease of the particle radius from $r^*_{debris}$ results in a gradual increase of the impact time. If $h \ge 100$ km, then the larger and largest particles will hit the Earth last (Fig.~\ref{fig4}B, Supplementary Video Clip VC2). If $h$ = 100 km, the larger and largest as well as smaller and smallest particles will simultaneously collide with the Earth last (Fig.~\ref{fig4}C, Supplementary Video Clip VC3). If $h$ < 100 km, the smaller and smallest particles will fall to Earth last (Fig.~\ref{fig4}D, Supplementary Video Clip VC4).
\par
Fig.~\ref{fig6}B displays the impact velocity $v$ as a function of the radius $r_{debris}$ of space particles launched with tangential ($\alpha = 90^\circ)$ circular orbit velocity $v_0$ (see Table~\ref{tab1}) from different launching height $h$. $v$ depends only slightly on $h$ in such a way that at a given $r_{debris}$, the impact velocity $v$ increases with increasing $h$, especially for larger particle radii.
\par
Fig.~\ref{fig6}C shows the impact angle $\beta$ from the local vertical (perpendicular to the Earth's surface) as a function of the radius $r_{debris}$ of space particles launched with tangential ($\alpha = 90^\circ$) circular orbit velocity $v_0$ (see Table~\ref{tab1}) for different launching height $h$. $\beta$ is practically independent of $h$, and increases from $0^\circ$ to $85^\circ$ as $r_{debris}$ increases from 0.01 mm to 10 m. Spherical space particles composed of iron with $r_{debris}$ < 10 mm fall to Earth practically perpendicular ($\beta \approx 0^\circ$) to the surface.
\par
Fig.~\ref{fig7} shows the trajectories of spherical particles at 1 h 12 min 8 sec after the explosion of a space debris at height $h$ = 1000 km above the Earth's surface. In this simulation the initial velocity of all explosion fragments was $v_0$ = 7 km/s and the angle of the initial velocity vector changed from $\alpha = 0^\circ$ to $\alpha = 360^\circ$ with an increment $\Delta \alpha = 12^\circ$ (see Supplementary Video Clip VC7). This visualization is at scale; the region above the Earth's surface $(h \ge 0)$ is not magnified for illustrative purposes. Outside the atmosphere, all explosion fragments with different radii launched with the same initial velocity vector $\vec{v}_0(v_0,\alpha)$ have the same elliptical trajectory (colored by red in Fig.~\ref{fig7} and Supplementary Video Clip VC7), because only the Earth's gravity influences their motion. After entering the atmosphere, explosion fragments with different radii launched with the same initial velocity $\vec{v}_0(v_0,\alpha)$ have different ballistic trajectories (colored differently in Fig.~\ref{fig7} and Supplementary Video Clip VC7) because of the size-dependent air drag.
\par
Fig.~\ref{fig8} displays the conditions of the initial velocity $v_0$ and angle $\alpha$ of spherical particles (with radius $r_{debris}$ = 0.01 mm and 10 m) of a space debris exploded at height $h$ = 100, 400, 1000 and 36000 km above the Earth's surface. If the initial velocity vector $\vec{v}_0(v_0,\alpha)$ is in the black region of the $v_0-\alpha$ map of Fig.~\ref{fig8}, the explosion fragment falls to Earth within 50 days.
\par
Table~\ref{tab2} summarizes the mean impact time <t> from a simulated explosion of an object. The debris is approximated as spherical particles originating from a height $h$ above the Earth's surface. The components of each particle's velocity, $v_0$ the magnitude and $\alpha$ the direction of the initial velocity vector $\vec{v}_0(v_0,\alpha)$  of the particles are varied from $v_0$ = 0 to 23 km/s with an increment $\Delta v_0$ = 0.23 km/s and $\alpha = 0^\circ$ to $\alpha = 360^\circ$ with an increment $\Delta \alpha = 3.6^\circ$. After the simulated explosion event at a given height $h$, on average the smaller  ($r_{debris}$ = 0.01 mm) fragments fall later (<t> = 317, 313, 314, 546 min) to Earth than the larger ( $r_{debris}$ = 10 m) ones (135, 157, 143, 401 min). The average impact times <t> are practically the same (313-317 or 135-157 min) at $h$ = 100, 400 and 1000 km for a given particle radius ( $r_{debris}$ = 0.01 mm or 10 m), while <t> (546 or 401 min) is much larger at $h$ = 36000 km.

\section{Discussion}\label{sec4}

The first spherical space debris (diameter: 58 cm, mass: 83.6 kg) originated from Sputnik-1, the first (Soviet) artificial Earth satellite launched in October 1957. Its elliptic orbit around the Earth had a perigeum and an apogeum distance of 215 and 939 km, respectively with a period of 96.2 minutes. After 1440 revolutions, on 4 January 1958, after 92 days of its launch, it entered the Earth's atmosphere and burned up (\url{https://en.wikipedia.org/wiki/Sputnik_1}).
\par
A space debris fragment can cause trouble in three ways: it (i) hits a spacecraft or satellite, or (ii) falls to Earth without fully burning up, or (iii) hits other debris and increases the amount of debris that continue to do damage. ~\citet{ESA2019} estimated the total number of space debris objects in orbit around Earth to be approximately 34 000 for sizes larger than 10 cm, 900 000 for sizes between 1 and 10 cm, and 128 000 000 for sizes from 1 mm to 1 cm. Most of the small debris pieces burn up totally in the atmosphere, but larger objects can reach the ground. The spherical pellets to be launched in the atmosphere from a high altitude in 2020 to produce an artificial meteor shower ~\citep{Greshko2016} will fully burn up, thus they will not fall to Earth.
\par
Space debris are presently rarely a concern for humans on Earth, because any space objects that do not burn up in the atmosphere are likely to fall into oceans, which cover over 70 \% of the Earth's surface or sparsely populated land areas. Due to the increasing number of space debris fragments, researchers are still looking for a solution that makes sure space debris burns up in the atmosphere. This solution of space debris reduction is strongly influenced by size and shape of the debris. The big and streamline-shaped debris are least likely to burn up. There are methods to change the shape of space debris. ~\citet{Monogarov2017} studied analytically and experimentally how to reduce the risk of such debris from impacting Earth. They used passive heating of the deorbited low-Earth orbit satellites and thus the so destructed satellite parts lost their streamline shape.
\par
Although the shape of space debris fragments can be arbitrary, in our present work we modelled the space mechanics of spherical debris particles only. Otherwise, the aerodynamical computations for any non-spherical particles would be extremely difficult. A possibly tumbling space debris fragment entering the denser atmosphere layers heats up due to air drag ~\citep{Allen1953} and smaller pieces may get off it. Because of the small direction changes of the fragment induced by these detachments, it is impossible to predict the temporal variations of its shape and rotation/tumbling. Thus, its exact trajectory, impact site and time cannot be computed. The simulation of trajectories of various non-spherical debris fragments with shapes of practical importance is a very difficult task ~\citep{Klinkrad2006}. Our results obtained for a spherical particle with radius $r_{debris}$ and density $\rho$ having area-to-mass ratio $Q_{spherical}$ = 3/($r_{debris}\rho$) can approximately be valid for a tumbling non-spherical space object with the same average Q-value: $Q_{spherical}$ = average $(Q_{non-spherical})$. It is reasonable to approximate tumbling debris as spherical, and thus it is possible to get valid results with less complex equations. In our simulations the shape factor (drag coefficient) was $c$ = 0.4 belonging to the sphere. A continuously and stochastically/randomly rotating/wobbling aspherical debris particle can be characterized by an effective/average shape factor. If this factor differs from that of the sphere (0.4), then the motion trajectories of such particles obviously differ from that of a sphere. In a future study it would be worth investigating the effect of different shape factors on the results presented here.
\par
There was at least one historic source of spherical particles of space debris with diameters exceeding 1 cm ~\citep{Klinkrad2006a}: 16 Russian RORSAT reactors released NaK droplets between 1980 and 1988. After the RORSAT transfer stage with the attached nuclear reactor has reached its near-circular disposal orbit of mean altitude 900-950 km, the reactor core was ejected. During ejection, most of the eutectic, liquid NaK alloy coolant of the primary cooling loop was released into space in the form of NaK spheres with density $\rho_{NaK} \approx 900$ kg/$\mathrm{m}^3$. With decreasing sizes, the NaK droplets encountered large perturbations due to radiation pressure and air drag, leading to short orbital lifetimes. After termination of NaK release events in 1988 the small-sized droplets were rapidly removed. As a consequence of increasing air drag during a solar activity peak in 1990 the sub-millimeter droplets decayed completely by 1992, and the sub-centimeter population was reduced by about 70 \%. Although, in our present study we neglected the solar radiation pressure, our results (Fig.~\ref{fig4}, Fig.~\ref{fig5}, Fig.~\ref{fig6}) are consistent with these empirical facts.
\par
The so-called 'Westford Needles' were deployed in two radio communication experiments in 1961 and 1963, when millions of copper needles (length 1.78 cm, diameter 17.8-25.4 $\mathrm{\mu}$m) were dispensed in two different orbits. Most of these needles cumulated into clusters. The single needles and small clusters had short orbit resident times due to their large area-to-mass ratios, and due to the corresponding level of radiation pressure and air drag perturbations ~\citep{Klinkrad2006a}. These observations also corroborate our results  (Fig.~\ref{fig4}, Fig.~\ref{fig5}, Fig.~\ref{fig6}).
\par
We did not take into consideration the burning up of spherical space particles in the atmosphere due to the extreme complexity of this physicochemical process ~\citep{Klinkrad2006}. For this task a real (non-static and non-exponential) atmosphere model - including the altitude-dependent chemical composition of the atmosphere - would be necessary. Due to burn, melt and evaporation, the shape and mass of a debris fragment continuously changes and decreases, respectively ~\citep{Klinkrad2006}, thus its drag coefficient changes also ~\citep{Marshall1966}. The simulation and experimental investigation of this complex process was out of the scope of our present work. Such computations have been performed in the prediction of break-up and survival of spacecrafts and satellites (e.g. ~\citet{Allen1953,Bouslog1994,Fritsche1999,Fritsche2000,Koppenwallner2001,Klinkrad2004,Rochelle2004,Klinkrad2006}.
\par
Many ideas have been developed to reduce the amount of space debris:
\par
(1) Satellites in low-Earth orbits can be disposed of by forcing them to re-enter the atmosphere, and most satellites in the less heavily trafficked geostationary region can be safely placed in 'graveyard' orbits that never interact with other objects. But in medium-Earth orbits (2000 km <$ h$ < 35000 km), satellite orbits can be unstable over long term because of gravitational resonances with the Moon and the Sun ~\citep{Witze2018}.
\par
(2) ~\citet{Daquin2016} showed that there is a dense web of orbital resonances that dictates how objects behave in medium-Earth orbits. There are paths in this web of resonances that lead not to medium-Earth orbits, but directly into the atmosphere. Thus, operators could take advantage of them to send unused satellites (space debris) straight into the atmosphere to burn up. This process is called 'passive disposal through resonances and instabilities'. Changing the launch date or time by as little as 15 minutes could lead to huge differences in how long a satellite remains in orbit. Such information could be used to help calculate the best times to depart the launch pad ~\citep{Daquin2016}. 
\par
	(3) Bigger debris pieces could be equipped with a sail ~\citep{Stohlman2013}. Such debris would slow down quicker and burn up in the atmosphere faster than without sail.
\par
	(4) A space debris fragment shot by a high-energy, well-concentrated, pulsing, ground-born LASER beam on its front surface, starts to evaporate, which results in its slowing down, the consequence of which is its height loss ~\citep{Esmiller2014}. Finally, the LASER-shot debris piece reaches the Earth's atmosphere where it burns up.
\par
	(5) The first active space RemoveDEBRIS technology was led by the Surrey Space Centre on 16 September 2018 (\url{https://www.sstl.co.uk/media-hub/latest-news/2018/removedebris-space-junk-net-capture-success}). At this occasion, the SpaceX Falcon 9 rocket passed a more than 100 kg heavy satellite to the International Space Station, which has laid a net on its trial projectile (\url{https://www.cnbc.com/2018/04/06/spacex-launches-removedebris-to-test-ways-to-remove-space-junk.html}, \url{https://qz.com/1395755/watch-this-space-robot-use-a-net-to-capture-errant-satellites/}).
\par
	(6) Satellites could be equipped with a new vision-based navigation which would be able to directly observe and lead specific, dangerous space debris ~\citep{Klinkrad2006a}.
\par
	(7) Satellites could be also supplied with a harpoon which would be able to place the captured space debris object at a lower orbit ~\citep{Klinkrad2010}.
\par
	(8) In order to avoid the transformation of the RemoveDEBRIS satellite into space debris, the satellite could be equipped with a large-surface drag-sail composed of a Helium-filled balloon ~\citep{Forshaw2016}. This balloon only opens at the end of the satellite's life cycle, then it slows down the satellite, thus it sinks to lower and lower orbits, finally it burns up in the atmosphere.
\par
Based on the above, we can see that one of the most obvious solutions of the space debris problem is to direct debris into the Earth's atmosphere where they can burn up. Therefore, it is important to examine how space debris fragments move in the atmosphere. In this work we gave an overall view about the typical trajectories of spherical space particles. We investigated the motion of these particles launched from both outside and inside the Earth's atmosphere. We computed the time, velocity and angle of impact as functions of the launch height, direction, speed and size of particles. All these can be important for both professionals and laymen and can also be used for meteoroids of the interplanetary dust entering the Earth's atmosphere.

\subsection*{Conflict of interest}

The authors declare no potential conflict of interests.

\subsection*{Ethics Statement}
 For our studies no permission, licence or approval was necessary.

\subsection*{Competing Interests}
The authors have no competing interests.

\subsection*{Authors' Contributions}

Substantial contributions to conception and design: JSB, GH\\
Software development: JSB\\
Performing computer simulations: JSB\\
Data visualization: JSB, DH, GH\\
Data analysis and interpretation: JSB, GH\\
Drafting the article or revising it critically for important intellectual content: JSB, RS, GH\\

\subsection*{Acknowledgements}

We are grateful to Mikl\'os Sl\'{i}z (software engineer, Graphisoft, Budapest, Hungary) for his help in the development of the computer simulation software. We thank Edward Brown (Graphisoft SE, Budapest) for improving the English of this paper.

\subsection*{Funding Statement}
There was no funding.

\section*{Supporting information}

The following supporting information is available as part of the online article (\url{https://onlinelibrary.wiley.com/doi/10.1002/asna.202023688}):

\noindent
\textbf{Video Clip VC1.}
{Positions (colored dots) of 1000 spherical iron (with density of $7.9\cdot10^3 \mathrm{kg/m^3}$) particles with different radius $r_{debris}$ until 2 h 18 min 14 sec after their launch with tangential ($\alpha=90^\circ$) circular orbit velocity $v_0 = 7.847$ km/s from the same point at height $h$ = 100 km from the Earth's surface (see  Fig.~\ref{fig4}A). The region above the Earth's surface ($h \ge 0$) is 100 times magnified for illustrative purposes. Different radii $r_{debris}$ are coded with different colors from violet for the smallest (0.01 mm) to red for the largest ($10^4$ mm) particle.}
\par
\noindent
\textbf{Video Clip VC2.}
{Trajectories (series of dots of a given color) of 100 iron particles with different radius $r_{debris}$ until 6 h 19 min 17 sec  after their launch with tangential ($\alpha=90^\circ$) circular orbit velocity $v_0 = 7.817 $ km/s from height $h$ = 150 km (see  Fig.~\ref{fig4}B). The region above the Earth's surface is 66 times magnified. Different radii $r_{debris}$ are coded with different colors from violet for the smallest (0.01 mm) to red for the largest ($10^4$ mm) particle.}
\par
\noindent
\textbf{Video Clip VC3.}
{Trajectories of 20 iron particles with different radius $r_{debris}$ until 2 h 10 min 27 sec after their launch with tangential ($\alpha=90^\circ$) circular orbit velocity $v_0 = 7.877 $ km/s from height $h$ = 50 km (see Fig.~\ref{fig4}C). The region above the Earth's surface is 200 times magnified. Different radii $r_{debris}$ are coded with different colors from violet for the smallest (0.01 mm) to red for the largest ($10^4$ mm) particle.}
\par
\noindent
\textbf{Video Clip VC4.}
{Trajectories of 100 iron particles with different radius $r_{debris}$ until 2 h 18 min 14 sec after their launch with tangential ($\alpha=90^\circ$) circular orbit velocity $v_0 = 7.847  $ km/s from height $h$ = 100 km (see Fig.~\ref{fig4}D). The region above the Earth's surface is 100 times magnified. Different radii $r_{debris}$ are coded with different colors from violet for the smallest (0.01 mm) to red for the largest ($10^4$ mm) particle.}

\noindent
\textbf{Video Clip VC5.}
{Trajectories of 100 iron particles with different radius $r_{debris}$ until 6 h 44 min 38  after their launch with initial velocity $v_0 = 4.57$ km/s and angle $\alpha=122.6^\circ$ relative to the radial direction from the same point at height $h$ = 10000 km above the Earth's surface (see Fig.~\ref{fig5}A). The region above the Earth's surface is not magnified.  Different radii $r_{debris}$ are coded with different colors from violet for the smallest (0.01 mm) to red for the largest ($10^4$ mm) particle.}

\noindent
\textbf{Video Clip VC6.}
{Trajectories of 100 iron particles with different radius $r_{debris}$ until 2 h 31 min 29 sec  after their launch with initial velocity velocity $v_0 = 5.65 $ km/s and angle $\alpha=45^\circ$  from  height $h$ = 100 km above the Earth's surface (see Fig.~\ref{fig5}B).The region above the Earth's surface is 100 times magnified.  Different radii $r_{debris}$ are coded with different colors from violet for the smallest (0.01 mm) to red for the largest ($10^4$ mm) particle.}

\noindent
\textbf{Video Clip VC7.}
{ Trajectories of 30 (different angles $\alpha$) $\times$ 11 (different radii $r$) = 330 spherical iron particles until 1 h 54 min 20 sec after the explosion of a space debris at height $h$ = 1000 km above the Earth's surface (see Fig.~\ref{fig7}). The initial velocity of all explosion fragments was $v_0$ = 7 km/s and the angle of the initial velocity vector changed from $\alpha = 0^\circ$ to $\alpha = 360^\circ$  with an increment $\Delta \alpha = 12^\circ$ (see Supplementary Video Clip VC7) where the region above the Earth's surface is not magnified.  Different radii $r_{debris}$ are coded with different colors from violet for the smallest (0.01 mm) to red for the largest ($10^4$ mm) particle.}

\noindent
\textbf{Gravitational potential of the geoid.}
{ }

\nocite{*}
\bibliography{Wiley-ASNA}%

\end{document}